\documentclass[a4paper,12pt]{article}

\usepackage{amssymb}
\usepackage{amsmath}
\usepackage{amsfonts}
\usepackage{amssymb}
\usepackage{graphicx}
\usepackage{color}%
\usepackage[utf8]{inputenc}
\usepackage{color}%
\usepackage[utf8]{inputenc}

\newtheorem{lemma}{Lemma}

 \newtheorem{remark}{Remark}
\newtheorem{theorem}{Theorem}

\newcommand{\FY}{F_0} 
\newcommand{\EE}{\mathbb E}
\newcommand{\PP}{\mathbb P}
\newcommand{\pilim}{\pi_{\infty}}
\newcommand{\glim}{g_{\infty}}
\newcommand{\Flim}{F_{\infty}}
\newcommand{\bX}{\mathbf{X}}
\newcommand{\bbeta}{\boldsymbol{\beta}}
\newcommand{\btheta}{\boldsymbol{\theta}}

\newcommand{\bgamma}{\boldsymbol{\gamma}}

\newcommand{\II}{\mathrm I}

\newcommand{\transpuesta}{\tiny{t}} 
\newcommand{\robusto}{\text{\tiny{R}} }

\newcommand{\gY}{\II_{ \{Y\leq y\}}} 

\newcommand{\SY}{\tiny{\hbox{SY}}}

\newcommand{\DPS}{\tiny{\hbox{DP-S}}}
\newcommand{\DPP}{\tiny{\hbox{DP-G}}}
\newcommand{\error}{\tiny{\hbox{error}}}
\newcommand{\glm}{\tiny{\hbox{GLM}}}
\newcommand{\Gglm}{\tiny{\hbox{G}}}
\newcommand{\reg}{\tiny{\hbox{REG}}}
\newcommand{\ipw}{\tiny{\hbox{IPW}}}

\newcommand{\DPPR}{\tiny{\hbox{DP-G-ROB}}}

\newcommand{\DPnor}{\tiny{\hbox{DP-NOR}}}
\newcommand{\DPnorR}{\tiny{\hbox{DP-S-ROB}}}

\title{ Robust Doubly Protected Estimators for Quantiles with Missing Data. }
\author{Julieta Molina$^{(1)}$, Mariela Sued$^{(2)}$, Marina Valdora$^{(1)}$, V\'ictor Yohai$^{(2)}$\ \\{\small (1) Universidad de Buenos Aires, (2) Universidad de Buenos Aires and Conicet.}}

\begin{document}

\maketitle
\begin{abstract}
Doubly protected methods are widely used for estimating the population mean of an outcome $Y$ from a sample
where the response is missing in some individuals. To compensate for the missing responses,   a vector $\bX$ of covariates is observed at each individual,
and the missing mechanism is assumed to be independent of the response, conditioned on $\bX$ (missing at random).  In recent years, many authors have turned from  the estimation of  the mean to that of the median, and  more generally,
doubly protected estimators of the quantiles have been proposed, under   a parametric regression model for the relationship between $\mathbf X$ and $Y$  and   a parametric form for the propensity score.
In this work, we present doubly protected estimators for
the quantiles that are also robust, in the sense that they are resistant
to the presence of outliers in the sample.  We also flexibilize the model for the relationship between $\mathbf X$ and $Y$.
Thus we present  robust doubly protected estimators for the quantiles of
the response in the presence of missing observations,
postulating a  semiparametric regression model for the relationship  between the response and the covariates and a parametric
model for the propensity score.
\end{abstract}

%
%


\maketitle


\section{Introduction}

The problem of estimating  the mean of a random variable $Y$
from an incomplete data set under the missing at random (MAR) assumption has attracted the attention of the statistical community during the last decades.  MAR
establishes that the variable   of interest $Y$ and the response indicator $A$ are
conditionally independent given an always observed vector $\textbf{X}$ of covariates.
Most of the existing proposals  are  based on three different approaches : inverse probability weighted (IPW), outcome regression (OR) and doubly protected (DP) methods.\\
Inverse probability weighted methods are based on the estimation of the  propensity score, denoted by $\pi(\textbf{X}),$
which is defined as the response
probability given $\textbf{X}$.   The estimators obtained using this  methodology are, essentially, weighted means of observed responses. The weights are determined as the inverse of the estimated propensity scores and their aim is to compensate for the missing observations.
Different approaches have been considered for the estimation of the propensity score used in the construction of IPW estimators, for example,
logistic regression models  \cite{littlerubin87, {kangshafer07}} and  splines \cite{littlean04}. \\
Outcome regression methods  require the estimation of the regression function  $g(\bX)=\EE(Y\mid \bX)$.
The estimators for $\EE(Y)$ built  by these techniques average predicted values computed using the estimated  regression function.   
Methods  to estimate  $g(\bX)$ include linear
regression \cite{yates1933}, kernel smoothing \cite{cheng1994},
semiparametric estimation \cite{wang2004,robinsetal95}  and  local polynomials \cite{gonza}.\\
IPW procedures  are consistent  for $\EE(Y)$ whenever the propensity score is properly estimated. For instance, in the parametric setting, this
requires a correctly specified   model
for $\pi(\bX)$.
Similarly,   OR  estimators are consistent
provided  the predicted values are based on a consistent estimation of the regression function. In a parametric framework, this means that
the postulated  model for the regression function $g(\bX)$ should be correct.
Doubly protected estimators, also known as doubly robust,  combine IPW and  OR methods  providing  consistent estimators for $\EE(Y)$  when either the model for the propensity score or
the model for the regression function is correct, without having to specify in advance
which of them holds. Thus, we get consistent estimators in the union of both the model
for the propensity and the model for the regression function.
In-depth analysis and examples of doubly protected methods are given in \cite{kangshafer07},\cite{lunceforddavidian04}, \cite{carpenter}, \cite{bangh}, and \cite{rob2003}.\\
Besides the estimation of $\EE(Y)$, many  authors have recently begun to apply these methods to the estimation of the quantiles of the distribution of $Y$ in the described MAR context.
Many  of them use the available techniques just described  to estimate the distribution function  $\FY$
of   $Y$ by $\widehat F_n$.
Then,  the  quantiles of $\widehat F_n$  are used to estimate those  of $\FY$.
Cheng and Chu \cite{chengchu96}, proposed a Nadaraya-Watson estimator  for the conditional distribution of $Y$ given $\bX$ and used it to derive a non parametric estimator for the distribution function $\FY$.
Yang, Kim and Shin \cite{yang2013} presented an imputation method for estimating the quantiles. Imputed data sets are constructed generating  plausible values  to represent the uncertainty about missing responses. The final quantile estimator is obtained combining  those computed at each  of the {multiple} imputed data sets. Wang and Qin (\cite{wang2010})  proposed to estimate  $\FY$ using IPW techniques but estimating the propensity score  via kernel regression. D\'iaz I. \cite{diaz2017} proposed to estimate $\FY$ under a semiparametric model, using doubly protected techniques by targeted maximum likelihood procedures.

Besides  the estimation of the quantiles of $\FY$, other location parameters have been considered and {some} of them also deal with the presence of outliers in the sample, providing robust methods for estimating the parameter of interest. For instance,   Bianco et al. \cite{bianco2010} used IPW techniques, imputing  non parametric estimators of the propensity score $\pi(\bX)$, and also regression procedures under a semiparametric partially linear regression model to construct location estimators. Asymptotic properties of the estimators involved in these proposals are presented in Bianco et al. \cite{bianco2011}.
Sued and Yohai \cite{suedyohai13}  also deal with the estimation of the entire distribution of $Y$   considering  a semiparametric regression model. Predicted values are combined with observed residuals to emulate a complete data set, based on which one can compute any desired estimator. A robust fit of the regression model is used to take care of anomalous observations. 
This proposal
consistently estimates  any parameter defined through a  weakly continuous functional at the response's distribution.


Causal inference is an area where missing data inevitably occurs because counterfactual variables
 may never be observed simultaneously. A large amount of procedures have been designed in such a framework.
Among them, we can cite the work by Lunceford and Davidian \cite{lunceforddavidian04} and the work by Zhang et al. \cite{zhang} on quantile estimation. Lunceford and Davidian proposed modified IPW estimators in the causal setting, achieving a higher precision compared to the classical ones. Zhang et al. presented several proposals for estimating the distribution function $\FY$, all of them based on parametric models for the propensity score or for the conditional distribution of $Y$ given $\bX$. They  also constructed doubly protected estimators for $\FY$ and, therefore, for the quantiles. No protection is provided against the effect of outliers, while the regression framework is entirely parametric.

All the doubly protected proposals to estimate the distribution function $\FY$ when $Y$ is missing in some individuals  considered up to now are very sensitive to anomalous observations.
This is due to the fact that they are based on least squares fits and/or in standard maximum likelihood techniques.
In this work, we introduce resistant doubly protected  estimators for the distribution function of a scalar outcome that is missing by happenstance on some individuals under a parametric model for the propensity score and a semiparametric regression model  for the relation between the outcome and the covariates, assuming missing at random responses.

The paper is organized as follows. Section \ref{sec:missing} introduces the classical procedures to estimate the mean of an outcome $Y$ missing at random.  In  Section \ref{section-Fy} we adapt these procedures to the estimation of $\FY$, the distribution function of $Y$ and briefly discuss on the importance of robustness in statistics. 
In Section \ref{seq:consistency} we establish the required  conditions to get a doubly protected estimator for the percentiles and  
 present a robust doubly protected estimator of the $p$- quantiles of $\FY$.  
In order to 
assess the performance of our estimating procedure, in Section \ref{MC} we present the results of a Monte Carlo simulation study. In Section \ref{example}, a real data set is used to compare the performance of many of the existing techniques. The proofs are presented in the Appendix.

\section{Estimating the mean of $Y$ .}
\label{sec:missing}

{Let us start considering} the problem of estimating  $\mu=\EE(Y)$
based on a sample $(X_1, A_1, Y_1),$ $ \ldots, (X_n,A_n,  Y_n)$,  distributed as $(X, A, Y)$,  where
$Y$ is missing by
happenstance on some subjects.
To compensate for the missing responses, a vector of covariates $\bX\in \mathbb{R}^p $ is available for each individual. Moreover, we assume
 that data are missing at random (MAR) \cite{rubin76}.
MAR establishes  that the  missing mechanism is not related to the response of interest and is only
related to the observed vector of covariates.
To give a formal definition, let $A$ be the indicator of
whether or not $Y$ is missing, i.e., $A=1$ if $Y$ is observed, and $A=0$ when $Y$ is missing.
Mathematically, MAR establishes that
$ \PP(A=1\mid  \boldsymbol X,Y)=\PP(A=1\mid \boldsymbol X)=\pi(\bX).$
$\pi(\boldsymbol X)$ is known in the literature as the propensity score \cite{rosenbaumrubin83}.

\subsection{ Inverse probability weighted  estimators. }
\label{sub1}
{Under the described framework,} $\EE(Y)$  can be represented in terms of the distribution of the
observed data as $\EE(Y)=\EE\left\{AY/\pi(\textbf{X})\right\}$. This representation 
{motivates the so-called  Horvitz and Thompson  \cite{horvitzthompson52} estimator of $\mu$}, defined by
\begin{equation}
 \label{ipw_missing}
 \widehat\mu(\widehat \pi) =\mathbb P_n\left\{\frac{AY}{\widehat \pi_n(\textbf{X})}\right\},
\end{equation}
where $\widehat\pi_n(\textbf{X})$ is a consistent estimator of $\pi(\textbf{X})$, and $\mathbb P_n $ is the empirical mean operator, i.e.
$\mathbb P_n V=n^{-1}\sum_{i=1}^n V_i.$
Notice   that the observed responses are weighted according to the inverse of the estimated probability  of $A=1$  given $\bX$,
justifying the appellative ``inverse probability weighted'' (IPW) for such procedures.
Moreover, those observed responses corresponding to  low values of the estimated propensity score are highly weighted since they should compensate for the high missing rate associated to such a level of covariates.
 For more details see \cite{robinsrotnitzky92}.

Different proposals to estimate $\pi(\textbf{X})$  give rise to diverse  estimators of $\EE(Y)$,
according to (\ref{ipw_missing}).
Nonparametric estimators of the propensity score have been considered by
Little and An \cite{littlean04}  and,  in a causal context, by   Hirano, Imbens and Ridder \cite{hiranoetal03}.

 In order to assume MAR,  the vector $\textbf{X}$ is   typically   high dimensional,  and therefore, in practice, non parametrical estimation of the propensity score is infeasible because of the so-called curse of dimensionality (see  \cite{stone1980}). For this reason the propensity score is often estimated postulating a parametric  working model,
assuming that
\begin{equation}
\label{modelpropen}
\pi(\textbf{X})=\pi(\textbf{X};\boldsymbol{\gamma}_0),
\end{equation}
where $\boldsymbol{\gamma}_0  \in  \mathbb{R}^q $   is an unknown $q$-dimensional parameter and $\pi(\cdot;\cdot): \mathbb R^p \times \mathbb R^q\to [0,1]$  is a known function.
The maximum-likelihood (ML) estimator  of $\boldsymbol{\gamma}_0$ can be obtained from $(\bX_1, A_1),\ldots, (\bX_n, A_n)$, and will be denoted by
$\widehat \bgamma_n$. Then, we can estimate $\pi(\bX)$ by 
\begin{equation}\label{pin}
\widehat\pi_n(\bX):=\pi(\bX; \widehat\bgamma_n).
\end{equation}

Generalized linear models (GLM) are very popular in this setting.
These models postulate that
$\pi(\textbf{X};\boldsymbol{\gamma}_0) =\phi(\boldsymbol{\gamma}_0^{\transpuesta} \textbf{X})$,
where  $\phi$ is a strictly increasing cumulative distribution function. In particular,
the linear logistic regression model for the propensity score is obtained by
choosing  $\phi(u) = exp(u)/\{1 +exp(u)\}$ (see \cite{kangshafer07} ).

\subsection{Outcome regression estimators.}
\label{sub2}
To construct IPW estimators, we model the relation between the missing mechanism and
the covariates and we do not make   assumptions on the
relation between the outcome and the covariates.
To construct regression estimators for $\EE(Y)$,
we estimate the regression function of $Y$ on $\bX$ and average predicted values.  More precisely,
under the missing at random  assumption we have that $Y\mid\bX \sim Y\mid (\bX; A=1)$ and therefore the regression function $g(\bX)=\EE\left(Y \mid \bX\right)$  satisfies
$g(\bX)=\EE\left(Y \mid \bX;A =1\right)$. In this way we arrive at a second representation for $\EE(Y)$ based on  the observed data through the regression function:  $\EE(Y)=\EE\{g(\bX)\}$.
 This alternative characterization  for $\EE(Y)$  invites us to estimate it {by} averaging  predicted values:
\begin{equation}
\label{reg_missing}
\widehat \mu(\widehat g_n) = \mathbb P_n\left\{ \widehat g_n(\mathbf{X}) \right\},
\end{equation}
where   $\widehat g_n(\textbf{X})$ is any consistent estimator of $g(\textbf{X}).$
Different ways to estimate $g(\bX)$ result in different   estimators according to  (\ref{reg_missing}).
 A non parametric proposal for $\widehat g_{n}(X)$ is given  in  Cheng \cite{cheng1994}, using  kernel regression estimation
duly adapted to the missing data context, in order to estimate the regression function. Imbens Newey and Ridder \cite{imbensetal05} proposed  non parametric estimation for both the propensity score and the regression function.

In practice, working  parametric models are postulated for the regression function to overcome the curse of dimensionality, which is a serious obstacle for non parametric methods (see  \cite{stone1980}).
A parametric model for the regression function assumes that
\begin{equation}
 \label{modelreg}
g(\textbf{X})=\;g(\bX;\bbeta_0),
 \end{equation}
 where $g(\cdot;\cdot)$ is a known function and
 $\boldsymbol{\beta}_0 \in \mathbb{R}^r  $ is unknown. The unknown parameter $\bbeta_0$ of the working regression model (\ref{modelreg}) can be estimated by $\widehat {\bbeta}_n$, using the units with observed responses $Y$ by, for instance, the least squares method. Then,   $g(\bX)$ is estimated with 
\begin{equation} \label{eq:estg}
\widehat {g}_n(\bX):=g(\bX; \widehat \bbeta _n)
 \end{equation} 
   and this expression is imputed in (\ref{reg_missing})
to obtain parametric regression estimators of $\EE(Y)$.

Kang and Shafer \cite{kangshafer07} review the regression estimator of $\mu$ that results  from considering a linear model for the regression function: $g(\bX; \bbeta)={\bbeta}^{\transpuesta} \bX$.
A  comprehensive overview of parametric regression estimators is given in \cite{littlerubin87}.

We also want to mention that another way to deal with the curse of dimensionality is to consider intermediate structures like additive models
or semiparametric models for the regression function. See, for example, \cite{wang2004} and \cite{hastie1990}.

\subsection{Doubly protected estimators.}
\label{sub3}
Estimators based on inverse probability weighting, as presented in (\ref{ipw_missing}) are consistent for $\mu$  as far as the propensity score $\pi(\bX)$ is consistently estimated. This approach leads to a well specified model for $\pi(\bX)$  in the parametric case.
  On the other hand, regression estimators, as presented in (\ref{reg_missing}), are consistent when  the  regression function is properly estimated and thus, the  regression model is assumed to be correctly specified.

To sum up,  each procedure forces us to choose in advance what to model in order to decide which estimator should be used to  consistently estimate  $\EE(Y)$.
 Typically no one knows which model is more suitable, generating  a  debate  on  which approach should be used. To end this controversy,  estimators which are  
 consistent  for $\EE(Y)$
 whenever, at least, one of the two models is correct were proposed. Such estimators  confer more protection to model misspecification than  IPW or OR estimators, which are consistent only when the corresponding assumed model holds. Since these estimators are   consistent for $\EE(Y)$ as long as one of
  the models  succeeds, they  are   called doubly protected estimators.

%

 Doubly protected estimators   were  discovered  by
 Robins et al. \cite{robinsetal94,robinsetal95},
while  studying   augmented  IPW estimators (AIPW).
Some years later, Scharfstein et al.  \cite{schar1999} showed that some AIPW  estimators  have the double protection  property. To motivate these estimators,  they obtained the following expression for the mean that holds  assuming  MAR and that  either $p(\textbf{X})=\PP(A=1\mid \textbf{X})$ or $r(\textbf{X})=\EE(Y\mid \textbf{X})$ hold:
\begin{equation}
\label{doblepobla}
\mu=\EE\left\{\frac{AY}{p( \textbf{X})}\right\}
-\EE\left[\left\{ \frac{A}{p(\textbf{X})}-1\right\} r(\textbf{X})\right].
\end{equation}
{Therefore, doubly protected estimators can be obtained by  postulating parametric 
models $\pi(\bX;\bgamma_0)=\PP(A=1\mid \bX)$ and $g(\bX;\bbeta_0)=\EE(Y\mid \bX)$, like in  (\ref{modelpropen}) and \eqref{modelreg}, respectively, and
 estimating $\mu$ by}
\begin{equation}
\label{equadoblec}
\widehat \mu_{DP}=
\mathbb P_n \left\{\frac{AY}{\widehat\pi_n(\textbf{X})}\right\}
-\mathbb P_n \left[\left\{ \frac{A}{\widehat\pi_n(\textbf{X})}-1\right\} \widehat g_n(\textbf{X})\right],
\end{equation}
where  $\widehat{\pi}_n$ and $\widehat g_n$ are defined in (\ref{pin}) and \eqref{eq:estg}.

The estimator defined in (\ref{equadoblec}) is doubly protected,  and also achieves full efficiency in the AIPW  class if the model for the propensity and the model for the regression function are both well specified (see \cite{robinsetal94} and \cite{robinsetal95}). An in-depth analysis and examples of doubly-protected methods are given in \cite{lunceforddavidian04}, \cite{carpenter}, \cite{bangh} and \cite{kangshafer07}. For the complete and detailed mathematical theory underlying double protection methodology see \cite{rob2001} and \cite{rob2003}.

\section{Estimating {the quantiles} of the distribution of $Y$}
\label{section-Fy}
 In this section  we move from the estimation of $\mathbb E(Y)$ to the estimation of the median of  the {distribution function} of $Y$. More generally, we will focus on the estimation of any quantile of $F_0$,  the {distribution function} of $Y$. The $p$-quantile of a distribution $F$ is defined as
 \begin{equation}
 T_p(F)=\inf\{x: F(x)\geq p\} \label{qt}.
  \end{equation}

{ When $p=0.5$,
$T_{0.5}(F)$ is the median   of $F$.}
The representation of the  $p$-quantile given in (\ref{qt}) suggests that it can be estimated by  $T_p(\widehat F_n)$,
provided $\widehat F_n$ approximates $\FY$. So, to estimate a $p$-quantile of a distribution of $Y$  it is enough
to estimate $\FY.$
Note that
$\FY\left(y \right)=\mathbb P(Y\leq y)=\mathbb E\left(\gY\right)$.
In the next sections, we will provide different proposals of  estimation  for the distribution function of $Y$,  mimicking each of the estimators considered in the previous sections, but to estimate now $\mathbb E\left\{\ell_y(Y)\right\}$, where $\ell_y(Y)= \II_{\{Y\leq  y\}}$, in lieu of $\mu=\EE(Y)$.

\subsection{IPW estimators of $\FY$. }

Under {the MAR}  assumption
\begin{equation}
\label{Fmar}
\FY(y)=\mathbb E\left\{\II _{\{Y\leq  y\}}\right\}=
\mathbb E
\left\{\frac{\,A\gY}{\pi(\mathbf{X}) }\right\},
\end{equation}
and so, if we apply the procedure developed in Section \ref{sub1},
we arrive at the following estimator for  $\FY(y)$
\begin{equation}
\label{ipw_dist}
 \widehat F_{\ipw}\left(y\right) = \mathbb P_n \left\{\frac{\,A \II_{\{Y\leq  y\}}}{\widehat{\pi}_n(\mathbf{X}) }\right\},
\end{equation}
where $\widehat{\pi}_n(\mathbf{X})$  is a consistent estimator of the propensity score.
A slightly modified version of this estimator, with weights adding up to one, has already been introduced
by Bianco et al. \cite{bianco2010} and used for estimating any M-location functional
of the distribution of $Y$. $ \widehat F_{\mbox{ipw}}$  was also proposed by Zhang et al. \cite{zhang} for estimating quantiles under a parametric model for the propensity score.

\subsection{Regression Estimators of $\FY$}
\label{sec-Fyreg}


Regression estimators are constructed based on the following representation
\begin{equation}
\FY(y)=\EE\{\PP(Y\leq y\mid \bX)\}.
\end{equation}
 Thus, we can estimate $\FY$ by
\begin{equation}
\label{reg}
 \widehat F_{\reg}(y)=\PP_n \widehat{\PP}(Y\leq y\mid \bX)=\frac{1}{n}\sum_{i=1}^n \widehat{\PP}(Y\leq y\mid \bX_i).
\end{equation}
Suppose, for instance, that a generalized linear model is postulated, assuming that
$Y\mid \bX\sim G_{{\bbeta_0}^t\bX}$, where $\{G_{\btheta}: \btheta\in \Theta\}$ is an exponential
 family of univariate distributions, and  $\btheta$ is the vector of natural parameters.
In this case, because of the MAR assumption, $Y\mid (\bX; A=1)\sim G_{\bbeta_0^{\transpuesta}\bX}$ and so
$\bbeta_0$ can be consistently estimated  with
$\widehat\bbeta_n$, the maximum likelihood estimator under the model
using the pairs $(\bX_i, Y_i)$ with $A_i=1$. Then, according to (\ref{reg}), in this particular case, we estimate $\FY$ with
\begin{equation}
 \label{glm}
 \widehat F_{\glm}(y)=\frac{1}{n}\sum_{i=1}^n G_{\widehat\bbeta_n^{\transpuesta}\bX_i}(y).
\end{equation}

Assume now that $Y$ follows a regression model of the form
\begin{equation}
\label{regression}
 Y=g(\mathbf{X})+u,
\end{equation}
where $g$ is the  unknown regression function that  maps $\mathbb R^p$ into $\mathbb R$, and  $u$ is  independent of $\mathbf {X}$. Moreover,  to guarantee the MAR assumption we require { $(\mathbf{X}, A)$ to be}  independent of $u$.  Let $\widehat g_n(\bX)$ be a consistent estimator of the regression function,
constructed using  the pairs $(\mathbf{X}_i, Y_i)$ with observed responses ($A_i=1$).
Under the regression model  (\ref{regression}), the assumed independence between $u$ and $\bX$ implies that
$\PP(Y\leq y\mid \bX)=\PP\{u\leq y-g(\bX)\}$ and so,
it can be estimated with the empirical distribution of the residuals $\widehat u_j=Y_j-\widehat g_n(\bX_j)$, for $A_j=1$,   at $ y-\widehat{g}_n(\bX)$:
\begin{equation}
\label{Ferror}
\widehat\PP(Y\leq y\mid \bX)= \widehat F_{\error}( y-\widehat g_n(\bX))=\frac{1}{m}\sum_{j=1}^n A_j \II_{\{\widehat u_j\leq   y-\widehat g_n(\bX)\}},
\end{equation}
where $m$ denotes the number of observed responses, that is $m=\sum_{j=1}^n A_j$.
Combining (\ref{Ferror}) with (\ref{reg}) we obtain the following estimator for $\FY$

\begin{equation*}
\widehat F(y)=
\frac{1}{n}\sum_{i=1}^n \widehat F_{\error}( y-\widehat g_n(\bX_i))
=\frac{1}{nm}\sum_{i,j=1}^n A_j  \II_{\{\widehat u_j\leq   y-\widehat g_n(\bX_i)\}}.
\end{equation*}

Note that $\widehat F$ assigns mass $1/(nm)$ to the values
\begin{equation}
\label{pseudoSY}
\widehat Y_{ij}= \widehat{g}_n(\mathbf{X}_i)+\widehat u_j\;, \quad 1\leq i,j\leq n,\,j;\, A_j=1,
\end{equation}
where predicted values $\widehat{g}_n(\mathbf{X}_i)$ are combined with residuals $\widehat u_j$ to emulate a pseudo-sample of responses $\widehat Y_{ij}$, as suggested by equation (\ref{regression}).

Sued and Yohai \cite{suedyohai13},  proposed a semiparametric regression model for (\ref{regression}), where
the regression function is assumed to be in a parametric family: $g(\mathbf{X})=g(\mathbf{X};\bbeta_0)$, with $\bbeta_0 \in  B\subset \mathbb R^q$, and $g: \mathbb R^p\times B \to \mathbb R$ is a known function.   No other than a centrality condition, namely symmetry around zero, is imposed on the error term $u$.
In fact, Sued and Yohai \cite{suedyohai13} showed that the centrality condition  can be avoided, redefining properly the intercept
in the regression model. But, to keep this presentation more accessible, we {can} focus on the centered
error case. This gives rise to the estimator $\widetilde  F$, defined by 
\begin{equation*}
\widetilde  F(y)=\frac{1}{nm}\sum_{i,j=1}^n A_i  \II_{\{\widehat Y_{ij}\leq y\}}, \end{equation*}
where  $\widehat Y_{ij}$ are defined as in (\ref{pseudoSY}) with $\widehat g_n(\bX)=g(\bX;\widehat \bbeta_n)$.
The authors proved that $\widetilde  F$
converges to $\FY$, as far as $\widehat{\bbeta}_n$ converges to $\beta_0$.
In particular, this procedure allows the estimation of the $p$-quantiles of $\FY$ with $T_p(\widetilde  F)$, where  $T_p$ is  defined in (\ref{qt}).

Model (\ref{regression}) with a linear regression  function $g(\bX; \bbeta)=\bbeta^{\transpuesta}\bX$ and
Gaussian errors $(u\sim \mathcal N(0, \sigma^2))$
fits the GLM framework described at the beginning of this section. In particular, according to (\ref{glm}), we arrive at the following estimator of $\FY$
\begin{equation}
 \widehat F_{\Gglm}(y)= \frac{1}{n}\sum_{i=1}^n \Phi\left( \frac{y-
\widehat{ \boldsymbol{\beta}}_n^{\transpuesta}\bX_i}{\widehat{\sigma}}
\right),
\end{equation}
where $\Phi$ denotes the cumulative distribution of  a standard normal random variable. This estimator was studied by Zhang et al. in \cite{zhang}.
We remark the semiparametric nature of the $\widehat F_{\SY}$ presented in \eqref{FySY}, where no model is assumed for the distribution of the error term $u$.

\subsection{Doubly protected estimators of $\FY$}

Replacing $Y$ with $\gY$ in (\ref{doblepobla}), we obtain
\begin{equation}
 \label{DPF}
\FY(y)=\EE\left\{\gY\right\}=
\EE\left\{\frac{A\gY}{p(\bX)}\right\}
-\EE\left[\left\{ \frac{A}{p(\bX)}-1\right\} r_y(\bX)\right],
 \end{equation}
 under MAR, and supposing either $p(\bX)=\PP(A=1\mid \bX)$ or  $r_y(\bX)=\PP(Y\leq y\mid \bX)$ {holds}.
Let $\pi(\bX)=\PP(A=1\mid \bX)$ and $g_y(\bX)=\PP(Y\leq y\mid \bX)$.  $\FY$ can be {doubly protectedly estimated } through a plug - in procedure inspired in  expression \eqref{DPF}, by
\begin{equation}
\label{DP-nopara}
\widehat F(y)=\frac{1}{n}\sum_{i=1}^n \frac{A_i\II_{\{Y_i\leq y\}} }{\widehat \pi_n(\bX_i)}
- \frac{1}{n}\sum_{i=1}^n \left\{ \frac{A_i}{\widehat \pi_n(\bX_i)}-1\right\} \widehat g_y (\bX_i).
\end{equation}

{At this point, we turn to a parametric framework to model  the propensity score with $\pi(\bX)=\pi(\bX;\boldsymbol{\gamma}_0)$.  
We also assume that model \eqref{regression} holds and that  the regression function $g(\bX)$  satisfies $g(\bX)=g(\bX;\bbeta_0)$.}
Thus, we can deal with a semiparametic regression model, instead of a parametric one, where typically also the distribution of the error term is assumed to belong to a parametric family.
In this way, using
$\widehat F_{\error}$ defined in (\ref{Ferror}),  we arrive at the following semiparametric  doubly protected  estimator for $\FY$
%

\begin{eqnarray}
 \label{DPFsemi}
\widehat F_{\DPS}(y)=\frac{1}{n}\sum_{i=1}^n \frac{A_i\II_{\{Y_i\leq y\}} }{\pi(\bX_i; \widehat {\boldsymbol{\gamma}}_n )}
- \frac{1}{mn}\sum_{i,j=1}^n \left\{ \frac{A_i}{ \pi(\bX_i; \widehat {\boldsymbol{\gamma}}_n)}-1\right\} A_j
\II_{\{g(\bX_i;\widehat \bbeta_n)+\widehat u _j \leq y
\}}  ,
\end{eqnarray}
where  $\widehat {\boldsymbol{\gamma}}_n$ and $\widehat {\boldsymbol{\beta}}_n$  are   estimators of $\boldsymbol{\gamma}_0$  and
$\boldsymbol{\beta}_0$, respectively.

{The GLM presented  in Section \ref{sec-Fyreg} can also be used to impute an estimator of   $g_y(\bX) =\mathbb P(Y\leq y\mid \bX)$ in the formula given in \eqref{DP-nopara}. For instance,  the linear model with  Gaussian errors ($u\sim \mathcal N(0, \sigma^2)$) is a particular case that  was already considered in \cite{zhang}, giving rise to the following formula for estimating $\FY$:  }

\begin{equation}\widehat F_{\DPP}(y)=\frac{1}{n}\sum_{i=1}^n \frac{A_i\II_{\{Y_i\leq y\}} }{\pi(\bX_i; \widehat {\boldsymbol{\gamma}}_n )}
- \frac{1}{n}\sum_{i=1}^n \left\{ \frac{A_i}{ \pi(\bX_i; \widehat {\boldsymbol{\gamma}}_n)}-1\right\}
\Phi\left( \frac{y-
\widehat{ \boldsymbol{\beta}}_n^{\transpuesta}\bX_i}{\widehat{\sigma}}
\right),
\label{defDPP}
\end{equation}
where $\Phi$ denotes the cumulative distribution of  a standard normal random variable, {$\widehat{\beta}_ n$ is the  least square estimator of the regression {coefficients},  and $\widehat \sigma$ estimates the standard deviation of the errors. }

\subsection{Robustness}\label{sec:rob}
Atypical observations,   called outliers,   are common in many real datasets. Classical procedures do not contemplate their existence and therefore  their  application  may lead to wrong conclusions.
For instance, the sample mean or the least-squares fit of a regression model, can be very adversely influenced by outliers, even by a single one. Robust methods arise to cope with these atypical observations, mitigating their impact in the final analysis.
The median  is, probably,
the most popular example of a robust procedure  to summarize a {univariate dataset.} More generally, M-location estimators have been developed for such a purpose: a robust location summary of  a univariate data set. Thus, moving from the estimation of $\EE(Y)$
to that of the median of $\FY$ represents a first step in the path towards robustification.
However, the methods presented in the previous section, also require regression fits,
both to estimate the propensity score and the regression function.
In this work we propose to consider  a robust alternative  for the  regression fit
of the postulated model that  relates $\bX$ and $Y$. In fact, when model  (\ref{modelreg}) is combined with a parametric regression function  assuming that
$g(\bX)=g(\bX; \bbeta_0)$,  the  least-squares fit  will be replaced by a robust one. 
{One way to achieve this robustification is to replace
the square loss function by a 
so called $\rho$-function 
 evaluated at the norm of  standardized  residuals.   A  $\rho$-function,   $\rho: \mathbb R\to [0,\infty]$, 
 is assumed to be  (i) continuous, (ii) even,  (iii)  non-decreasing function of $|t|$ and (iv) $\rho(0)=0$. Moreover, in order to deal with high leverage outliers the $\rho$-function should be bounded (see Section 5.4.1 in \cite{maronnamartinyohai06} ).} 
 
This is the case of  M-estimators, both for location and regression problems.  In particular, MM-estimators, introduced by Yohai \cite{yohai1987} for the linear model, and extended to the non linear case in  Fasano et al.
\cite{Fasano2012}, combines the highest possible tolerance to the presence of outliers, measured by the breakdown point, with an arbitrarily high efficiency
in the case of Gaussian errors.

{The use of a robust fit for the regression model to estimate $\FY$ has already been considered in \cite{suedyohai13}. Indeed, they   presented $\widehat F_{SY}$,  a semiparametric regression estimators  as those  discussed in Section \ref{sec-Fyreg},  where  the regression parameter $\bbeta_0$  is estimated with $\widehat \bbeta^{\robusto }_n$, an MM- estimator. Thus,}
\begin{equation}
\label{FySY}
\widehat F_{SY}(y)=\frac{1}{nm}\sum_{i,j=1}^n A_i  \II_{\{g(\bX_j;\widehat \bbeta^{\robusto}_n)+Y_i-g(\bX_i;\widehat \bbeta^{\robusto}_n)\leq y\}}.  \end{equation}
The authors showed that $T(\widehat F_{SY})$ gives rise to a robust method for estimating $T(\FY)$, for any weak-continuous functional $T$ at $\FY$.

{Robust doubly protected estimations of $\FY$ can be obtained {by} replacing    the least squares estimator of $\bbeta_0$  by  $\widehat\bbeta^{\robusto}_n$, an MM-estimator. For instance, a robust version  of the doubly protected estimator discussed in \cite{zhang}, presented in \eqref{defDPP}, can be defined as }   
\begin{equation}\widehat F_{\DPPR}(y)=\frac{1}{n}\sum_{i=1}^n \frac{A_i\II_{\{Y_i\leq y\}} }{\pi(\bX_i; \widehat {\boldsymbol{\gamma}}_n )}
- \frac{1}{n}\sum_{i=1}^n \left\{ \frac{A_i}{ \pi(\bX_i; \widehat {\boldsymbol{\gamma}}_n)}-1\right\}
\Phi\left( \frac{y-
\bX_i^{\transpuesta}{\widehat \bbeta^{\robusto }_n}}{\widehat{s}}
\right),
\label{DPPROB}
\end{equation}
{where $\widehat s$ is a robust scale of the residuals.}
 
{
An MM-estimator can also be used to robustify the semiparametric doubly protected estimator $\widehat F_{\DPS}$, 
defined in \eqref{DPFsemi}. We  postpone this construction until next section, where $\widehat F_{\DPS}$ is  
slightly modified in order get a  consistent procedure. In this way, we will be able to present  both a  classical and a robust 
 doubly protected  estimator of $\FY$.
}



%

The next step is, naturally, to robustify the estimation of $\bgamma_0$, the parameter involved in the propensity score. However, this simple approach would not robustify the final estimator. On the contrary, in the presence of outliers in the covariates, extreme values of $\widehat{\pi}_n$ are more likely to appear if $\bgamma_0$ is estimated robustly than  otherwise. Resistance to outliers in $A$ seems a difficult problem, whose solution we are still working on and might be the subject of further work.

\section{Consistency}\label{seq:consistency}
The estimator $\widehat F_{\DPS}(y)$ defined in equation (\ref{DPFsemi})
is not  a cumulative distribution function of a probability measure.
However,  it can be associated to a discrete
signed measure on $\mathbb R$. Moreover, we can decompose it as
\begin{equation}
\label{DP-nopara-expandido}
\widehat F_{\DPS}(y)=\frac{1}{n}\sum_{i=1}^n \frac{A_i\II_{\{Y_i\leq y\}} }{\widehat \pi_n(\bX_i)}
-\frac{1}{nm}\sum_{i,j=1}^n  \frac{A_i}{ \widehat \pi_n(\bX_i)} A_j
\II_{\{\widehat{g}_n(\bX_i)+\widehat u _j \leq y
\}}  +\frac{1}{nm}\sum_{i,j=1}^n A_j
\II_{\{\widehat{g}_n(\bX_i)+\widehat u _j \leq y
\}}, 
\end{equation}
with $\widehat\pi_n(\bX)=\pi(\bX;
\widehat\bgamma_n)$ and $\widehat g_n (\bX)=g(\bX;\widehat  \bbeta_n)$. Let $\widehat F_1(y)$, $\widehat F_2(y)$
and $\widehat F_3(y)$ denote the three terms in expression \eqref{DP-nopara-expandido}. Only the last one of them,  $\widehat F_3$,  corresponds to a cumulative distribution function  of a probability measure.
The total mass of neither the first  term, $\widehat F_1$,  nor the second one,  $\widehat F_2$,  is equal to one. This issue can be easily corrected  normalizing them properly.
Let $\widetilde F_1$ be the normalized correction of $\widehat F_1$, namely
\begin{equation}
\label{Cn}
\widetilde F_1:=\frac{1}{C_n}\sum_{i=1}^n \frac{A_i\delta_{Y_i} }{\widehat \pi_n(\bX_i)} \text{,  where  }   C_n :=\sum_{i=1}^n \frac{A_i }{\widehat \pi_n(\bX_i)}
,\\
\end{equation}
and $\delta_s$ denotes the distribution function of  the point mass probability  measure concentrated    at $s$.
To normalize the second term involved in expansion (\ref{DP-nopara-expandido}), consider
\begin{equation}
\widetilde F_{2a}:=\frac{1}{C_n}\sum_{i=1}^n \frac{A_i\delta_{\widehat g_n(\bX_i)} }{\widehat \pi_n(\bX_i)}
\;,\quad \widetilde G:=\frac{1}{m}\sum_{j=1}^n A_j \delta_{\widetilde u_j},\\
\label{F2tilde}
\end{equation}
and, therefore,  the normalized version of $\widehat F_2$, is given by
\begin{equation}
\label{tildeF2}
\widetilde F_2:=\widetilde F_{2a}\ast \widetilde G,
\end{equation}
where $\ast$ stands for the  convolution operator between two distribution functions.
Finally, recalling that $\widehat F_3$ is already a cumulative distribution function, note that it can be written as
\begin{equation}
\widehat F_3=\widehat F_{3a}\ast \widetilde G,
\quad \hbox{with }
\quad \widetilde F_{3a}=\frac{1}{n}\sum_{i=1}^n \delta_{\widehat g_n(\bX_i)}.
\label{F3}
\end{equation}
The normalized version of the doubly protected estimator presented in (\ref{DP-nopara-expandido})  is given by
\begin{equation}
\label{DP-normalized}
\widehat F_{\DPnor}=\widetilde F_1-\widetilde F_{2a}\ast \widetilde G +\widetilde F_{3a}\ast \widetilde G. 
\end{equation}
{
At this point we want to emphasize that, even though each term of the sum in (\ref{DP-normalized}) is a cumulative distribution function, the non convexity of the linear combination that defines $\widehat F_{\DPnor}$ causes it to fall out of the space of cumulative distribution functions. However, Theorem \ref{consiquantil} states that, under assumptions A1-A3 given below,  $\widehat F_{\DPnor}$ is a doubly protected estimator of $\FY$ since   it converges uniformly to $\FY$ almost surely (a.s.), if
either  $\widehat \pi_n(\bX)$  converges to $\pi(\bX)$ or $\widehat g_n(\bX)$ converges to  $g(\bX)$. Even though $\widehat F_{\DPnor}$ is not a cumulative distribution function, Lemma \ref{lemmaconsi} says that $T_p(\widehat F_{\DPnor})$ is well defined, with $T_p$ as in (\ref{qt}).
Finally, the aforementioned theorem states that
$T_p(\widehat F_{\DPnor})$ converges to $T_p(\FY)$ a.s.  if
either  $\widehat \pi_n(\bX)$  converges to $\pi(\bX)$ or $\widehat g_n(\bX)$ converges to  $g(\bX)$.  Therefore, $T_p(\widehat F_{\DPnor})$ results a  doubly robust estimator of the $p$-quantile of the distribution of $Y$, as established in Theorem \ref{consiquantil}.}

{In order to study the {asymptotic behaviour} of $\widehat F_{\DPnor}$, defined in (\ref{DP-normalized}), {we}
consider the following assumptions.} 

\begin{itemize}
\item{A1:} 
{There exists  a sequence of random functions $(\widehat \pi_n)_{n\geq 1}$,  $\widehat \pi_n: \mathbb R^p\to (0,1)$, depending on 
$(X_1, A_1), \ldots, (X_n, A_n)$, i.i.d., distributed, as $(X,A)$, such  that  $\sup_{\bX\in \mathcal S_{\bX}} \vert \widehat \pi_n(\bX)-\pilim(\bX)\vert \to 0$ a.s., for some function  $\pilim: \mathbb R^p \to (0,1)$, { where  $\mathcal S_{\bX}$ stands for 
 the support of the distribution of $\bX$.} }

\item{A2:} 
$\inf_{\bX\in \mathcal S_\bX} \pilim(\bX)=i_\infty>0$
\item{A3:} 
{There exists  a sequence of random functions $(\widehat g_n)_{n\geq 1}$,  $\widehat g_n: \mathbb R^p\to \mathbb R$, depending on $(X_1, A_1, Y_1),$ $\ldots, (X_n, A_n, Y_n)$, such that,  for every compact set $\mathcal K$, $\sup_{\bX\in \mathcal K} \vert \widehat g_n(\bX)-\glim(\bX)\vert \to 0$ a.s.,  for some function  $\glim: \mathbb R^p \to \mathbb R$.}

\end{itemize}
Let  $\phi=\mathbb E\{\pi(\bX)/\pilim(\bX)\}$, consider the distribution functions
\begin{eqnarray}
\label{F12}
&&F_1(y)=\frac{1}{\phi} \mathbb E\left\{\frac{\pi(\bX)}{\pilim(\bX)}\gY \right\}\;,
\quad
F_{2a}(y)=\frac{1}{\phi} \mathbb E\left\{\frac{\pi(\bX)}{\pilim(\bX)}\II_{ \{\glim(\bX)\leq y\}} \right\}\;,\\
\label{F3G}
&&F_{3a}(y)=F_{ \glim(\bX)}(y)
\;, \quad G(y)= F_{ \{Y-\glim(\bX)\}\mid A=1}(y)
\end{eqnarray}
and let
\begin{equation}
\label{F-lim}
\Flim=F_1-F_{2a}\ast G +F_{3a}\ast G.
\end{equation}
The following result  indicates in which circumstances $\Flim$ coincides with $\FY$.

\begin{theorem}
\label{bienlimite}
Assume that the propensity score   $\pi(\bX)=\mathbb P(A=1\mid \bX)$ is equal to $\pilim(\bX)$. Then, $F_1=\FY$, $F_{2a}=F_{3a}$ and therefore, $\Flim=\FY$.
Consider now the regression model   $Y=g(\bX)+u$, with $u$ independent of $(A, \bX)$, and assume that $g(\bX)=\glim(\bX)$.
Then, $F_1=F_{2a}\ast G$, $\FY=F_{3a}\ast G$
and consequently $\Flim=\FY$.
\end{theorem}

The next theorem establishes the double robustness of the quantile estimators.

\begin{theorem}
\label{consiquantil}
Assume that  $Y=g(\bX)+u,$  with $u$ independent of $(A, \bX),$ and let $\pi(\bX)$ denote the propensity score  $\pi(\bX)=\mathbb P(A=1\mid \bX)$ .{ Let $\{(X_i, A_i, Y_i)\}_{i\geq 1}$ be independent {and} identically distributed as $(X, A, Y)$. }
Assume that conditions A1-A3 are satisfied, that the cumulative distribution function $G$ of  $\{Y-\glim(\bX)\}\mid (A=1)$
is continuous.
Assume also that either $g(\bX) = \glim(\bX)$ or $\pi(\bX)=\pilim(\bX)$.
Then, 
\begin{equation}
\label{res1}
\sup_y\vert \widehat F_{\DPnor}(y)- \FY(y)\vert \to 0 \quad \text{a.s.}
\end{equation}

Moreover, $T_{ p}(\widehat F_{\DPnor})$ is well defined and  for every  $p\in (0,1)$ such that  $\FY$ is strictly increasing in a neighborhood of $T_{p}(\FY)$, we have 
 \begin{equation}
 \label{res2}
T_{ p}(\widehat F_{\DPnor})\to T_{p}(\FY) \quad\hbox{a.s., }
 \end{equation}
if $g(\bX) = \glim(\bX)$ or $\pi(\bX)=\pilim(\bX)$.
\end{theorem}

\subsection{Robust Doubly Protected Estimators for Quantiles.}
\label{seq:consistency-rob}
We will now combine the {robust notions} discussed  in Sections \ref{sec:rob} with the consistency result presented in Section  \ref{seq:consistency} to get a robust doubly protected estimator of the $p$-quantile  $T_p(\FY)$. To do so, {we consider  parametric models $\pi(\bX;\boldsymbol \gamma)$ and $g(\bX;\boldsymbol\beta) $ for the propensity score and the regression function respectively and assume that at least one of them holds.}
Let $\widehat{\pi}_n(\bX)=\pi(\bX;\widehat \gamma_n)$, where $\widehat \gamma_n$ is the MLE under the {postulated parametric} model for the propensity score.  The regression model will be fit with an MM- estimator, which will be denoted with $\widehat \bbeta_n^{\robusto}$, and therefore, we define   $\widehat g_n^{\robusto}(\bX)=g(\bX;\widehat \bbeta_n^\robusto ) $, while the residuals obtained from this procedure will be denoted with  $\widehat u^{\robusto}$; that is to say, $\widehat u_j^{\robusto}=Y_j-g(\bX_j;\widehat \bbeta_n^{\robusto} )$, for $j$ such that $A_j=1$. 
We propose to estimate the $p$- quantile $T_p(\FY)$ with $T_p(\widehat F_{\DPnorR})$, where $\widehat F_{\DPnorR}$ is  the following semiparametric  normalized robust doubly  protected estimator for $\FY$:

\begin{equation}\small
\label{eq:DPnorR_sued_2}
\widehat F_{\DPnorR}=\frac{1}{C_n}\sum_{i=1}^n \frac{A_i\delta_{Y_i} }{ \widehat \pi_n(\bX_i)}
-\frac{1}{C_nm}\sum_{i,j=1}^n  \frac{A_i}{ \widehat \pi_n(\bX_i)} A_j
\delta_{\widehat g_n^{\text{\tiny R}}(\bX_i)+\widehat u _j^{\text{\tiny R}} }  +\frac{1}{nm}\sum_{i,j=1}^n A_j
\delta_{\widehat g_n^{\text{\tiny R}}(\bX_i)+\widehat u _j^{\text{\tiny R}} },
\end{equation}
{The following theorem establishes that  $T_{ p}(\widehat F_{\DPnorR})$ is a   doubly protected   estimator of $T_{ p}(\FY)$. 
}

\begin{theorem}\label{teo:consrob-bis}
Assume that   $Y=g(\bX)+u,$  with $u$ independent of $(A, \bX)$. 
Let $\{(X_i, A_i, Y_i)\}_{i\geq 1}$ be an i.i.d. sequence, distributed as $(X, A, Y)$. Denote with 
  $\widehat \bgamma_n$  the MLE  assuming a logistic regression model $\phi(\bgamma^{\transpuesta}\bX)$ for the propensity score $\PP(A=1\mid \bX)$.  Let $\widehat\bbeta_n^{\robusto}$ be an MM-estimator, under the  linear model  $\bbeta^{\transpuesta}\bX$ for the regression function $g(\bX)$.  
    Assume that 
\begin{itemize}
\item[(i)] There exist $\bgamma_\infty$ and $\bbeta_\infty^R$ such that $\widehat \bgamma_n\to \bgamma_\infty$ a.s. and $\widehat{\bbeta}^{\robusto}_n\to \bbeta^{\robusto}_\infty$ a.s.
 \item[(ii)]   $\mathcal S_{\bX}$ is compact.
 \item[(iii)]
 The  cumulative distribution function $G$ of  $\{Y-\bX^{\transpuesta}\bbeta_\infty^{\robusto})\}\mid (A=1)$
is continuous.
\item[(iv)]  Either $\PP(A=1\mid \bX)=\phi(\bgamma_0^{\transpuesta}\bX)$, for some $\bgamma_0$, or $g(\bX)=\bbeta_0^{\transpuesta}\bX$, for some $\bbeta_0$.
\end{itemize}
 Then
$$T_{ p}(\widehat F_{\DPnorR})\to T_p(\FY)\quad a.s.$$
\end{theorem}

\begin{remark} Both MLE and MM estimators are particular cases of M- estimators and  therefore, under regularity conditions,  their  limit point can be characterized, regardless the validity of the assumed model.  In particular, under regularity conditions, the maximum likelihood estimator $\widehat{\boldsymbol\gamma}_n$  converges a.s. to 
$$\boldsymbol\gamma_\infty =\arg\max_{\boldsymbol{\gamma}}\mathbb E\left\{\log p(\mathbf X,A,\boldsymbol\gamma) \right\}, $$
where $p(\mathbf X,A,\boldsymbol\gamma)= \pi(\mathbf X;\boldsymbol\gamma)^A\{1- \pi(\mathbf X;\boldsymbol\gamma)\}^{1-A}$, whether or not the  postulated model for the propensity score is correctly specified.  Similarly, whether or not the regression model is correctly specified,  MM-estimators converge, under regularity conditions, to 
$$\boldsymbol\beta_\infty=\arg\min_{\boldsymbol\beta} \mathbb E\left(\rho \left[ \{Y-\mathbf \bX\boldsymbol \beta\}/\sigma_\infty\right]\mid A=1\right),$$ for some $\sigma_\infty >0$ (see Theorems 2 and 3 in \cite{Fasano2012}).
\end{remark}

\section{Monte Carlo Simulation}
\label{MC}
In this section we report the results of a Monte Carlo study we made in order to analyze   the performance of the different estimators of the median proposed in this work, as compared to some of the estimators that already exist in the literature. We consider different distributions of the error term $u$, namely standard normal and student distribution with one (Cauchy distribution) and three degrees of freedom.  We also investigate the robustness  of the proposed estimators, by contaminating the samples with $10\%$ of outliers.
We consider samples of $n$ i.i.d. random vectors   $(\mathbf X_i,A_i,Y_i)$ where $\textbf{X}_i=(1,X_{i1},X_{i2})$ is a bivariate standard normal random vector of covariates, that is,  $\mathbf X_i \sim \mathcal N(\mathbf 0,\mathbf I)$, $A_i$ is a binary variable   following a Bernoulli distribution with
\begin{equation*}
P(A=1\vert \textbf{X})=\pi(\textbf{X})= \mbox{expit}( (1,X_1,X_2) \boldsymbol{\gamma}_0  ),
 \end{equation*}
where  $\mbox{expit}(x) = {e^x }/{(1+e^x)}$, $x\in \mathbb R$ and  $\boldsymbol{\gamma}_0^{\transpuesta}=(0, 0.1,-1.1)$; the outcome $Y$ satisfies
\begin{equation*}
 Y =  (1,X_1,X_2) \boldsymbol{\beta}_0 + u,
 \end{equation*}
where $\boldsymbol{\beta}_0^{\transpuesta}=(0, -3 , 2)$ and $u$ is independent of $(\textbf{X},A)$.

We consider four different situations:
\begin{enumerate}
\item[S.1] Both the model for the propensity score and the model for the regression function are well specified. 
\item[S.2]  The model for the propensity score is well specified {while} the model for the regression function is misspecified. More precisely,  we fit an incorrect model for the regression function, using just the covariate $X_1,$ that is, the covariate $X_2$ is omitted.
\item[S.3]  The model for the propensity score is  misspecified {while} the model for the regression function is well specified. The misspecification consists in fitting a logistic regression model with only the covariate $X_1$, that is, the covariate $X_2$ is omitted.
\item[S.4]  Both models are misspecified by omitting the covariate $X_2$.
\end{enumerate}
{ 
For each case, we generate $Nrep=1000$  samples of size $n=100$ and we compute 5  estimators of $\eta=\text{med}(Y)$ by evaluating $T_{0.5}$ at different estimators of $\FY$: $\widehat F_{\ipw}$,
$\widehat F_{\SY}$,  $\widehat F_{\DPP}$,  
$\widehat F_{\DPPR}$ and $\widehat F_{\DPnorR}$.  
{ The definitions of these estimator can be found in equations (\ref{ipw_dist}), (\ref{FySY}), (\ref{defDPP}), (\ref{DPPROB}) and \eqref{eq:DPnorR_sued_2}}. Henceforth we will use the subscript of each {of them} to invoke the corresponding  procedure; for instance $\ipw$ refers to the estimator $T_{0.5}(\widehat F_{\ipw})$. 
In each of the situations contemplated in S.1-S.4,  $\widehat \bgamma_n$ denotes the MLE computed under the postulated model for the propensity score;  $\widehat \bbeta_n$ and $\widehat \bbeta_n^{\robusto}$  denote the least square estimator and an MM estimator for the proposed linear model for the regression function $g(\bX)$. { As in \cite{suedyohai13}, we take, as $\widehat \bbeta_n^{\robusto}$, an  
 MM-estimator,
with $\rho_0$ and $\rho_1$ in the Tukey bisquare family,  $k_0= 1.57$, $k_1=3.44$ and $\delta=0.5$.}
In  display (\ref{defDPP}) $\widehat \sigma$  stands for the {clasical} {unbiased} estimator of error standard deviation while   $\widehat s$, in (\ref{DPPROB}), {is an MM scale estimator of the regression residuals.}

{Empirical mean square errors are presented in Table \ref{tabla}, where  
 PS stands for propensity score model, OR stands for outcome regression model,  and the other three columns correspond to different distributions
{of} the error term $u$ in the linear model. } {These results show that the double protection property of $\DPP$,  $\DPPR$ and $\DPnorR$ also holds for finite samples. }

{In order to investigate the robustness of the estimators, we contaminate with outliers samples generated as above, but with normal errors. To this end,  $10\%$ of the observations are replaced }by outliers $(\mathbf X_0,A_0, Y_0)$ where
{ $\mathbf X_0=(1,2,0)$,}
$ \PP(A_0=1 | \mathbf X= \mathbf X_0) = \mbox{expit}( (1,2,0) {\boldsymbol{\gamma}}_0  )$ and 
\begin{equation}\label{grid}
Y_0\in\left\{ -100,-90,\dots,-20,-10,0,10,20,\dots ,90,100\right\}.
\end{equation}
Simulation results under contaminations  are summarized in Table \ref{tabla2} and Figures \ref{fig:poro} to \ref{fig:pmro}. 
In Table \ref{tabla2} we show the maximum mean square error under $10\%$ of contamination for values of $Y_0$ in the grid given in (\ref{grid}).
Also in Table  \ref{tabla2}, PS stands for propensity score model and OR for outcome regression model. 

These results show that, even though the median is already robust, the estimation of the regression coefficients by a robust method improves the performance of the estimators. This improvement is very important if the sample is contaminated with outliers but it is also noticeable when the sample has a heavy tailed distribution such as a Student or a Cauchy distribution.
On the other hand,  $\DPnorR$ gives better results than $\DPPR$ when the errors follow a Student or Cauchy distribution. This is due to the fact that the latter assumes normal errors while the former does not.\\

In figures \ref{fig:poro} to \ref{fig:pmro} we plot the mean square errors of the different doubly protected estimators as a function of the value of the outlying outcome $Y_0$.
These figures show that much robustness is gained by estimating the regression coefficients robustly, using an MM-estimator, instead of the least squares estimator. Both doubly protected robust estimators give good results for contaminated samples.  Note that, as expected  $\DPPR$ outperforms $\DPnorR$ when the regression model is correctly specified, while  $\DPnorR$ outperforms $\DPPR$ when it is not.



\begin{table}[ht]
\centering
\begin{tabular}{rrrrrr}
  \hline
Estimator &PS&OR& Normal errors & t3 errors & Cauchy errors \\
  \hline
\ipw&correct& & 0.381 & 0.424 & 0.689 \\
\ipw&incorrect& & 1.125 & 1.137 & 1.22\\
\SY&&correct & 0.206 & 0.233 & 0.339  \\
  \SY&&incorrect & 0.945 & 1.035 & 0.996 \\
  \DPnorR&correct&correct & 0.313 & 0.361 & 0.548 \\
  \DPnorR&correct&incorrect & 0.280 & 0.310 & 0.469  \\
\DPnorR&incorrect&correct & 0.683 & 0.528 & 0.841  \\
\DPnorR&incorrect&incorrect & 0.983 & 1.115 & 1.100  \\
  \DPP&correct&correct & 0.310 & 0.361 & 0.707  \\
  \DPP&correct&incorrect & 0.268 & 0.326 & 0.740  \\
{\DPP}&incorrect&correct & 0.712 & 0.570 & 0.733 \\
  \DPP&incorrect&incorrect & 0.982 & 1.104 & 1.063  \\
  \DPPR&correct&correct & 0.310 & 0.364 & 0.590 \\
{\DPPR}&correct&incorrect & 0.278 & 0.302 & 0.483  \\
 {\DPPR}&incorrect&correct & 0.682 & 0.537 & 0.942  \\
\DPPR&incorrect&incorrect & 0.976 & 1.088 & 1.088  \\
   \hline
\end{tabular}
\caption{Mean square errors for different scenarios {under the central model (without contaminations)}}
\label{tabla}
\end{table}

\begin{table}[ht]
\centering
\begin{tabular}{rrrrrrr}
  \hline
Estimator &PS&OR&  Max MSE \\
  \hline
\ipw&correct& &  0.885 \\
\ipw&incorrect& &  2.300\\
\SY&&correct &  1.641 \\
\SY&&incorrect &   4.301 \\
\DPnorR&correct&correct & 0.675 \\
\DPnorR&correct&incorrect &  0.907 \\
\DPnorR&incorrect&correct &  0.733 \\
\DPnorR&incorrect&incorrect &2.355 \\
\DPP&correct&correct & 1.036 \\
\DPP&correct&incorrect &   1.131 \\
\DPP&incorrect&correct &  2.706 \\
\DPP&incorrect&incorrect &  2.314 \\
\DPPR&correct&correct &  0.695 \\
\DPPR&correct&incorrect  &0.945 \\
\DPPR&incorrect&correct  &  0.681 \\
\DPPR&incorrect&incorrect  &2.314 \\
   \hline
\end{tabular}
\caption{Maximum mean squared error under $10\%$ of outlier contamination and the regression model with normal errors.}
\label{tabla2}
\end{table}

\section{Example: Hospital data.}
\label{example}
We consider a sample of 100 patients hospitalized in a Swiss hospital during 1999
for medical back problems. We study the relationship between the cost of stay (Cost,
in thousands of Swiss francs) and some explanatory variables that are available on administrative records: length of stay (LOS, in days), admission type (0 = planned; 1 = emergency),
insurance type (0 = regular; 1 = private), age (years), sex (0 = female; 1 = male),
discharge destination (1 = home; 0 = another health institution). This data set has been analyzed in \cite{marazziyohai04} and has  no missing values.
In order to study the performance of our proposed estimators, we artificially delete some of the responses and compute the estimators in the sample with missing values. We repeat this procedure 1000 times

In each replication we generate a sample of dichotomous variables  $A_1 \dots A_n$ according to the following mechanism:

$$ \ln\left(    \frac{P(A_i=1) }{1-P(A_i=1}      \right)=0.1* \mathbf{LOS}_i-1.1 .$$

The responses with corresponding $A_i=0$ are deleted from the sample and considered missing.
In this way, the proportion of missing responses is approximately $0.5$.

For each sample  we compute estimators of the median Cost of stay by  five methods: $\ipw$, $\reg$, $\DPP$, $\DPPR$ and $\DPnorR$. These estimates are compared with the median cost of stay computed with the entire sample, $\eta=9.69
$, as follows: let $\widehat\eta$ be one of the five estimators mentioned above, then we estimate the mean square error of  $\widehat\eta$ by

$$ MSE=\frac{1}{1000} \sum_{i=1}^{1000} \left(\widehat\eta_i-\eta\right)^2,$$
where $\widehat\eta_i$ is the value of $\widehat\eta$ at the $i-th$ sample.

An analysis of the linear regression fit with the complete data set shows that all six variables considered are relevant to predict {Cost} and that no transformations are necessary; 
for this reason we consider this the ``correct'' model, both for the PS and the OR. To compare the fit with the one obtained if either model is misspecified, we also consider ``incorrect'' models,  which include all six covariates, but {\tt LOS} is transformed to $\log${\tt LOS}.



The results are summarized in Table \ref{table:ex1}. This example suggests that both $\DPnorR$ and $\DPPR$ have a good performance in real data sets with missing values, with better results than $\DPP$. $\DPPR$ is slightly better when the OR model is correctly specified, while $\DPnorR$ is a somewhat more resistant to its misspecification.




\bigskip 
  
\textbf{Acknowledgement} 

The authors thank Dr. Alfio Marazzi for the data set in the example and Dr. Ana Bianco and Dr. Graciela Boente for helpful discussions.
This research was partially supported by Grant  \textsc{pict} 2014-0351 from \textsc{anpcyt} and Grants 20020150200110BA and 20020130100279BA from the Universidad de Buenos Aires at Buenos Aires, Argentina
\vspace*{-8pt}

\section{Appendix}

\textbf{Proof of Theorem \ref{bienlimite}: } 
If $\pi(\bX)=\pilim(\bX)$,  then $\pi(\bX)/\pilim(\bX)=1$ and $\phi=1$. Therefore,  $F_1=\FY$, $F_{2a}=F_{3a}$ and $\Flim=\FY$. 

If $Y=g(\bX)+u$, with $u$ independent of $(A, \bX)$ and $g(\bX)=\glim(\bX)$, then $F_{3a}$ is the distribution function of $g(\bX)$ and $G$ is the distribution function of $u$.  Therefore,  $F_{3a}\ast G$ is the distribution function of $g(\bX)+u=Y$, that is to say $F_{3a}\ast G=F_0$.
On the other hand, let $Z$ be a random variable, {independent of $u$,} with distribution function $F_{2a}$, then $F_{2a}\ast G$ is the distribution function of $Z+u$, which, by definition, is equal to
\begin{eqnarray*}
P(Z+u\leq y)&=P(Z\leq y-u)
=  \frac{1}{\phi} \mathbb E\left\{\frac{\pi(\bX)}{\pilim(\bX)}\II_{ \{g(\bX)\leq y-u\}} \right\} = \frac{1}{\phi} \mathbb E\left\{\frac{\pi(\bX)}{\pilim(\bX)}\II_{ \{g(\bX)+u\leq y \}} \right\}=F_1(y). \ \;
\end{eqnarray*} 
{Thus, $\Flim=\FY$ also in this case.}
$\square$
\bigskip

The following five lemmas  will be used to prove Theorem \ref{consiquantil}. 
{Recall that  $\widetilde F_{1}$ and  $\widetilde F_{2a}$, defined in \eqref{Cn} and \eqref{F2tilde}, respectively,  are indeed  random sequences of  cumulative distribution functions  based on sample of size $n$ (which we omit in the notation).  }

\begin{lemma}
\label{F1} Consider $\widetilde F_1$ and $F_1$, defined in \eqref{Cn} and \eqref{F12}, respectively. Under assumptions A1 and A2, it holds that $\widetilde F_1$ converges  to  $F_1$ uniformly, a.s., that is  
\begin{eqnarray*}
\mathbb P\left(\sup_y\vert \widetilde  F_1(y)- F_1(y)\vert \to 0 \right)=1
\end{eqnarray*}
\end{lemma}

\textbf{Proof: }
We show first that $C_n/n\to \phi$ a.s. To do so, note that we can write
\begin{equation}
\label{cnlim}
\frac{C_n}{n} =\frac{1}{n}\sum_{i=1}^n \left\{\frac{A_i }{\widehat \pi_n(\bX_i)}-\frac{A_i }{ \pi_\infty(\bX_i)}\right\}+\frac{1}{n}\sum_{i=1}^n\frac{A_i }{\pi_\infty(\bX_i)}.
\end{equation}
By the law of large numbers, the second term in \eqref{cnlim} converges a.s.  to
$$\mathbb E \left\{   \frac{A}{\pi_\infty(\bX)}  \right\}=\mathbb E\left\{  \frac{1}{\pi_\infty(\bX)} \mathbb E \left( \left. A \right|X  \right)\right\}=\mathbb E \left\{   \frac{\pi(\bX)}{\pi_\infty(\bX)}  \right\}=\phi.$$
It remains to prove that
the first term in \eqref{cnlim} converges to zero a.s. Now, under conditions A1 and A2, given $\varepsilon \in (0,1)$ there exists $n_0$ such that $\left| \pi_\infty(\bX)-\widehat \pi_n(\bX)\right| <\varepsilon i_\infty$ for all $n\geq n_0$, and therefore, $(1-\varepsilon)i_\infty \leq \widehat \pi_n(\bX)$ for such $n$, implying that
\begin{equation}\frac{1}{n}\sum_{i=1}^n A_i \frac{\left| \pi_\infty(\bX_i)-\widehat \pi_n(\bX_i) \right| }{\widehat \pi_n(\bX_i) \pi_\infty(\bX_i)}<\frac{1}{n}\frac{1}
{(1-\varepsilon)i_\infty^2}\sum_{i=1}^n A_i \left| \pi_\infty(\bX_i)-\widehat \pi_n(\bX_i) \right| <\frac{\varepsilon}{(1-\varepsilon)i_\infty} .
\label{pipilim}
\end{equation}
and then we obtain the announced  result.\\

Second, we prove that
\begin{equation}
\label{p1prop1}
\mathbb P\left\{ \lim_{n\to \infty}\sup_y \left \vert \widetilde F_1(y)-
\frac{1}{\phi}\frac{1}{n}\sum_{i=1}^n \frac{A_i\II_{\{Y_i\leq y\}} }{ \pilim(\bX_i)}
\right\vert =0
\right\}=1.
\end{equation}
To prove \eqref{p1prop1}, notice that
adding  and subtracting
$(n\phi)^{-1}\sum_{i=1}^n {{A_i\II_{\{Y_i\leq y\}} }/{ \widehat \pi_n(\bX_i)}}$,  we get 
\begin{equation}
 \left \vert \widetilde F_1(y)-
\frac{1}{\phi}\frac{1}{n}\sum_{i=1}^n {\frac{A_i\II_{\{Y_i\leq y\}} }{ \pilim(\bX_i)}} \right  \vert \leq
\vert\{C_{n}/n\}^{-1}-{\phi}^{-1}\vert\;C_{n}/n
+
\frac{1}{\phi n}\sum_{i=1}^n A_i\vert \widehat \pi(\bX_i)^{-1}- {\pilim(\bX_i)}^{-1}\vert.
\label{paso2}
\end{equation}
Neither of the two terms in  \eqref{paso2} depend on $y$ and  they both converge to zero under A1-A2; the convergence of the first {term follows from} the convergence of $C_n/n  $  to $\phi$ a.s., while the convergence of the second one has already been proved in \eqref{pipilim}. This proves  \eqref{p1prop1}.

 Finally, using arguments similar to those in the proof of the Glivenko-Cantelli theorem (see, for instance, Theorem 19.1 in
\cite{vandervart00}
), it can be shown that
\begin{equation}
\label{p3prop1}
\mathbb P\left\{ \lim_{n\to \infty}\sup_y \left \vert
\frac{1}{n}\sum_{i=1}^n \frac{A_i\II_{\{Y_i\leq y\}} }{\pilim(\bX_i)}
- \mathbb E\left\{\frac{A\II_{\{Y\leq y\}}}{\pilim(\bX_i)}
\right\}\right\vert =0
\right\}=1.
\end{equation}
The result follows combining (\ref{p1prop1}) and( \ref{p3prop1}). $\square$

\bigskip 
Henceforth we use $G_n \xrightarrow{w} G$ to denote weak convergence of cumulative distribution functions.

\begin{lemma}
\label{F2a}Consider  $\widetilde F_{2a}$ and $F_{2a}$, defined in  \eqref{F2tilde} and \eqref{F12}, respectively.  Under assumptions A1-A3, it holds that $\widetilde  F_{2a}$ converges weakly to $F_{2a}$ a.s., i.e.,
\begin{eqnarray}
\label{resp1}
\mathbb P\left (\widetilde  F_{2a} \xrightarrow{w}  F_{2a}\right)=1
\end{eqnarray}
\end{lemma}

\textbf{Proof: }
Let $\mathcal C_{\text{buc}}$ denotes the set of functions
 $f:\mathbb R\to \mathbb R$  bounded and uniformly continuous.
In order to prove the lemma  we will show that

\begin{equation}
\label{pdauncont}
\mathbb P\left ( \lim_{n\to \infty}   \int f d \widetilde F_{2a} = \int f d F_{2a}  , \, \forall f \in  \mathcal C_{\text{buc}}  \right)=1.
\end{equation}
Let 
$$  \widetilde F_3(y)= \frac{1}{C_n}\sum_{i=1}^n
\frac{A_i\delta_{ \widehat g_n(\bX_i)}(y)}
{ \pilim(\bX_i)}\,\, \mbox{and}\,\, \widetilde F_4(y)= \frac{1}{C_n}\sum_{i=1}^n
\frac{A_i\delta_{\glim(\bX_i)}(y)}
{\pilim(\bX_i)}.
$$
{
Note that both  $\widetilde F_3$ and $\widetilde F_4$ defined above,  are sequences of random functions; however we omit $n$ in the notation for simplicity.}

Fix $ f \in \mathcal C_{\text{buc}}$. Defining   $I_1(f) = \left \vert  \int f d \widetilde F_{2a} - \int f d \widetilde F_{3}  \right \vert,$  $I_2(f) = \left \vert  \int f d \widetilde F_{3} - \int f d \widetilde F_{4}  \right \vert,$
and $I_3(f) = \left \vert  \int f d \widetilde F_{4} - \int f d F_{2a}  \right \vert$, we get that  
\begin{eqnarray}
\label{dtriang}
\left \vert  \int f d \widetilde F_{2a} - \int f d F_{2a}  \right \vert  \leq I_1(f) + I_2(f)+I_3(f). 
\end{eqnarray}
{Let us now consider each of these three terms.}
Since $f$ is bounded, {using arguments similar to those} in the proof of Lemma \ref{F1}, we have that under A1 and A2 
\begin{equation}
\label{I}
\mathbb P\left (\lim_{n\to \infty} \left \vert  \int f d \widetilde F_{2a} - \int f d F_{3}  \right \vert = 0, \, \forall f \in  \mathcal C_{\text{buc}}  \right)=1.
\end{equation}
\newline
To  deal with $I_2(f)$  notice that  
\begin{eqnarray}\label{despacito1}
I_2(f)=\left\vert  \frac{1}{C_n}\sum_{i=1}^n
\frac{A_i f\{{\widehat g_n(\bX_i)\}}} 
{\pilim(\bX_i)} -  \frac{1}{C_n}\sum_{i=1}^n
\frac{A_i f\{{\glim(\bX_i)\}}}
{\pilim(\bX_i)}
\right\vert\leq \frac{n}{C_n} \frac{1}{ni_\infty} \sum_{i=1}^n
\left\vert
f\{{\widehat g_n(\bX_i)\}} -f\{{\glim(\bX_i)\}}\right\vert
\end{eqnarray}
{Since $f$ is uniformly continuous, given $\varepsilon>0$, there exists $\delta$ such that $\vert u_1-u_2\vert <\delta$ implies
$\vert f(u_1)-f(u_2)\vert<\varepsilon$. Take} $K$ large and consider the compact set $\mathcal K=\{\vert\vert \bX\vert\vert \leq K\}$. For $n$ large enough, invoking now A3, we  get  that
 $\sup_{\bX\in \mathcal K} \vert \widehat g_n(\bX)-\glim(\bX)\vert <\delta$ and therefore,
 the right hand side of \eqref{despacito1} is smaller than
\begin{eqnarray}\frac{n}{C_n}\left( \frac{\varepsilon}{i_\infty}+
\frac{1}{ni_\infty} \sum_{i=1}^n
2\vert\vert f\vert\vert _\infty I_{\{\vert\vert\bX_i\vert \vert >K\}}\right), \end{eqnarray}
which implies that 
\begin{equation}
\label{II}
\mathbb P\left (\lim_{n\to \infty} \left \vert  \int f d \widetilde F_{3} - \int f d \widehat F_{4}  \right \vert = 0, \, \forall f \in  \mathcal C_{\text{buc}}  \right)=1.
\end{equation} 

It remains to show that  
\begin{equation}
\label{III}
\mathbb P\left ( \lim_{n\to \infty}  \int f d  \widetilde F_{4} = \int f d F_{2a}, \, \forall f \in  \mathcal C_{\text{buc}}  \right)=1
\end{equation}
Notice that, as in Lemma \ref{F1}, {using arguments similar to those in  the proof of  the }Glivenko-Cantelli theorem,
 we have that 
\begin{eqnarray}\label{chachito13}
&&\mathbb P\left( \lim_{n\to \infty}\sup_y \left\vert
\frac{1}{n} \sum_{i=1}^n
\frac{A_i\delta_{\glim(\bX_i)}(y)}
{\pilim(\bX_i)} -\mathbb E\left\{\frac{A\II_{\{ \glim(\bX) \leq y\}} }{ \pilim(\bX)}
\right\}
\right\vert=0
\right)=1
\end{eqnarray}
and therefore
\begin{eqnarray}\label{chachito13bis}
&&\mathbb P\left( \lim_{n\to \infty} 
\frac{1}{\widetilde C_n}
\sum_{i=1}^n
\frac{A_i\delta_{\glim(\bX_i)}(y)}
{\pilim(\bX_i)} =\frac{1}{\phi}\mathbb E\left\{\frac{A\II_{\{ \glim(\bX) \leq y\}} }{ \pilim(\bX)}
\right\}\;,\forall y\in \mathbb R
\right)=1,
\end{eqnarray}
where $\widetilde C_n=\sum_{i=1}^n A_i/\pilim(X_i)$. 
 Both the sequence  as the  limit   function  presented in  \eqref{chachito13bis} are cumulative distribution functions. By the MAR assumption,
 \begin{equation}\label{eq:3deprop2}
 \frac{1}{\phi}\mathbb E\left\{\frac{A\II_{\{ \glim(\bX) \leq y\}} }{ \pilim(\bX)}
 \right\}=F_{2a}(y)
 \end{equation}
 and, therefore, \eqref{chachito13bis}
 implies that 
 \begin{eqnarray}\label{chachito13bis-bis}
 &&\mathbb P\left( \lim_{n\to \infty} 
 \frac{1}{\widetilde C_n}
 \sum_{i=1}^n
 \frac{A_i f(\glim(\bX_i))}
 {\pilim(\bX_i)} =\int f d F_{2a}, \, \forall f \in  \mathcal C_{\text{buc}}  
 \right)=1.
 \end{eqnarray} 
  Finally, since $\widetilde C_n/C_n\to 1$, 
  we conclude that \eqref{III} holds. 
%
%
%
%
%
%
%
%
The result stated in the lemma follows from combining  \eqref{dtriang},  \eqref{I}, \eqref{II} and \eqref{III}. $\square$

\bigskip

{The following lemma was proved in \cite{suedyohai13}, as a part of Theorem 1.}


\begin{lemma}\label{F3a}
 Consider $\widetilde  F_{3a}$ and $\widetilde  G$, defined in \eqref{F3} and \eqref{tildeF2},  $F_{3a}$ and $G$ defined in \eqref{F3G}.
Under assumption A3, $\widetilde  F_{3a}$ converges weakly to 
$F_{3a}$ a.s. and  also $\widetilde  G$ converges wakly to $G$ a.s., i.e, 
\begin{eqnarray*}
\mathbb P\left (\widetilde  F_{3a} \xrightarrow{w}  F_{3a}\right)=1
\quad\hbox{and}\quad
\mathbb P\left (\widetilde  G \xrightarrow{w}   G\right)=1.
\end{eqnarray*}
\end{lemma}




{
As announced in Section \ref{seq:consistency}, we will now show that 
the functional $T_p$, presented  in  (\ref{qt}), can be defined over  an enlarged family of functions, which includes cumulative distribution functions, preserving  its continuity.}

\begin{lemma} \label{lemmaconsi} Consider a distribution function $F:\mathbb R\to [0,1]$ and $p\in (0,1)$ such that there exists a unique value $y_p$ with $F(y_p)=p$, and so $T_p(F)=y_p$, for $T_p$ defined in (\ref{qt}) . {Let  $F_{n}:\mathbb{R}  \to \mathbb{R}, n\geq 1$, be a sequence of functions} such that
\begin{enumerate}
\item  $\lim_{y \to - \infty} F_{n}(y)=0$ and $\lim_{y \to + \infty} F_{n}(y)=1$.
\item  $F_{n}$ converges uniformly to $F$.
\end{enumerate}
Then $T_p$ can be defined at $F_n$ and
$$\lim_{n \to \infty} T_p(F_n)=T_p(F).
$$
\end{lemma}
\textbf{Proof:}
 Let $ A_{n,p}=\left\{ y \in \mathbb{R} : F_n(y) \geq p  \right\}.$ By the assumptions of the lemma,  $\lim_{y \to + \infty} F_{n}(y)=1$,
 and therefore,
$A_{n,p}$ is not empty. Since $\lim_{y \to - \infty} F_{n}(y)=0$ we conclude that  $A_{n,p}$ is  bounded from below, and therefore
$T_p(F_n)=\inf A_{n,p}$ is well defined.

Given $\varepsilon>0$, let

$$
\delta=\min\left \{\left(  F(y_{p}+\varepsilon)-F(y_{p}))/2\right)  ,\left(
F(y_{0})-F(y_{p}-\varepsilon)\right)  /2\right \} .
$$
By the assumptions of the lemma, $\delta>0$. Now, the uniform convergence of $F_n$ to $F$ guarantees that there exists $n_0$ such that
$$\sup_{y \in \mathbb R}\vert F_n(y)-F(y)\vert \leq \delta\;, \hbox{for all $n\geq n_0$.} $$
In particular, 

\begin{align}
\sup_{y<y_{p}-\varepsilon}F_{n}(y)<F(y_p-\epsilon)+\delta\leq F(y_p)-2\delta+\delta \leq p-\delta.\label{uno}%
\end{align}
and
\begin{align}
F_{n}(y_{p}+\varepsilon)  \geq F(y_{p}+\varepsilon)-\delta \geq F(y_{p})+2\delta-\delta=p+\delta>p.\label{dos}%
\end{align}
From (\ref{uno}) and (\ref{dos})  we conclude that, for all $n\geq
n_{0}$ we have $|y_{n}-y_p|\leq\delta,\ $and therefore,
$y_{n}\rightarrow y_{p}$ en $ $ This concludes the proof. $\square$

\bigskip 

\textbf{Proof of Theorem \ref{consiquantil}:}
The continuity of $G$ implies that $F_{2a}\ast G$ and $F_{3a}\ast G$ are both  continuous cumulative distribution functions. Since weak convergence to a continuous limit distribution function implies uniform convergence (see, for example, Lemma 2.11 in \cite{vandervart00}),
Lemmas  \ref{F2a} and \ref{F3a}  imply that  $\widetilde F_{2a}\ast \widetilde G$ and $\widetilde F_{3a}\ast\widetilde  G$ converge uniformly to  $F_{2a}\ast G$ and $F_{3a}\ast G$, respectively, a.s.

 Combining these results  with Theorem \ref{bienlimite},  we obtain \eqref{res1}.
  From Lemma \ref{lemmaconsi}, we  conclude  that $T_p(\widehat F_{\DPnor})$ is well defined. Moreover, Lemma \ref{lemmaconsi} and the uniform convergence proved bellow,  implies that $T_p(\widehat F_{\DPnor})$ converges to $T_p(\FY)$ a.s.
 $\square$

\bigskip 
{
\textbf{Proof of Theorem \ref{teo:consrob-bis}:}

We will show that  A1-A3 are satisfied, with $\widehat{\pi}_n(\bX)=\phi(\widehat \gamma_n^{\transpuesta}\bX)$, 
 $\pilim(\bX)=\phi(\bgamma_\infty^{\transpuesta}\bX)$, $\widehat g_n (\bX)=\bbeta_n^{\transpuesta} \bX$ and $\glim(\bX)= \bbeta_\infty^{\transpuesta}\bX$.
}
To prove A1, note that  
\begin{equation}
\label{expansion-pi}
\vert \widehat \pi_n(\bX)-\pilim(\bX)\vert=\left\vert \pi(\bX; \widehat{ \boldsymbol \gamma}_n)-\pi(\bX; \boldsymbol\gamma_\infty )\right\vert=\left\vert \phi^\prime ( \widetilde{\boldsymbol\gamma}_n^{\transpuesta}\bX) \bX^{\transpuesta} (\boldsymbol {\widehat{\boldsymbol\gamma}_n}- \boldsymbol \gamma_\infty)\right\vert,
\end{equation}
where $\widetilde{\boldsymbol\gamma}_n$ is an intermediate point between  $\boldsymbol {\widehat{\gamma}}_n$ and $\boldsymbol \gamma_\infty$. The convergence of $\widehat \bgamma_n$ to $\bgamma_\infty$ a.s. combined with the assumed compactness for the support of $\bX$ imply  the validity of A1.

A2 is satisfied since $\phi(\bgamma_\infty^{\transpuesta}\bX)$  is continuous and $\bX$ has a compact support. 

To prove the validity of A3, observe that 
$$\vert \widehat g_n(\bX)-\glim(\bX)\vert= \vert   \{\boldsymbol {\widehat{\beta}_n}-\bbeta_\infty\}^{\transpuesta}\bX\vert.$$
 The convergence of $\widehat \bbeta_n$ to $\bbeta_\infty$ a.s. guarantees that    A3 is also satisfied.
 
 Finally, note that if $\mathbb P(A=1\mid \bX)=\phi(\bgamma_0^{\transpuesta}\bX)$, then $\bgamma_\infty=\bgamma_0$, and so $\pilim(\bX)=\PP(A=1\mid X)$. Also, if $g(\bX)=\bbeta_0^{\transpuesta}\bX$, then $\bbeta_\infty=\bbeta_0$ implying that $\glim(\bX)=g(\bX)$.  We can now invoke Theorem \ref{consiquantil} to conclude the proof of the {theorem}.
  $\square$

\begin{figure}
\begin{center}\includegraphics[scale=0.5]{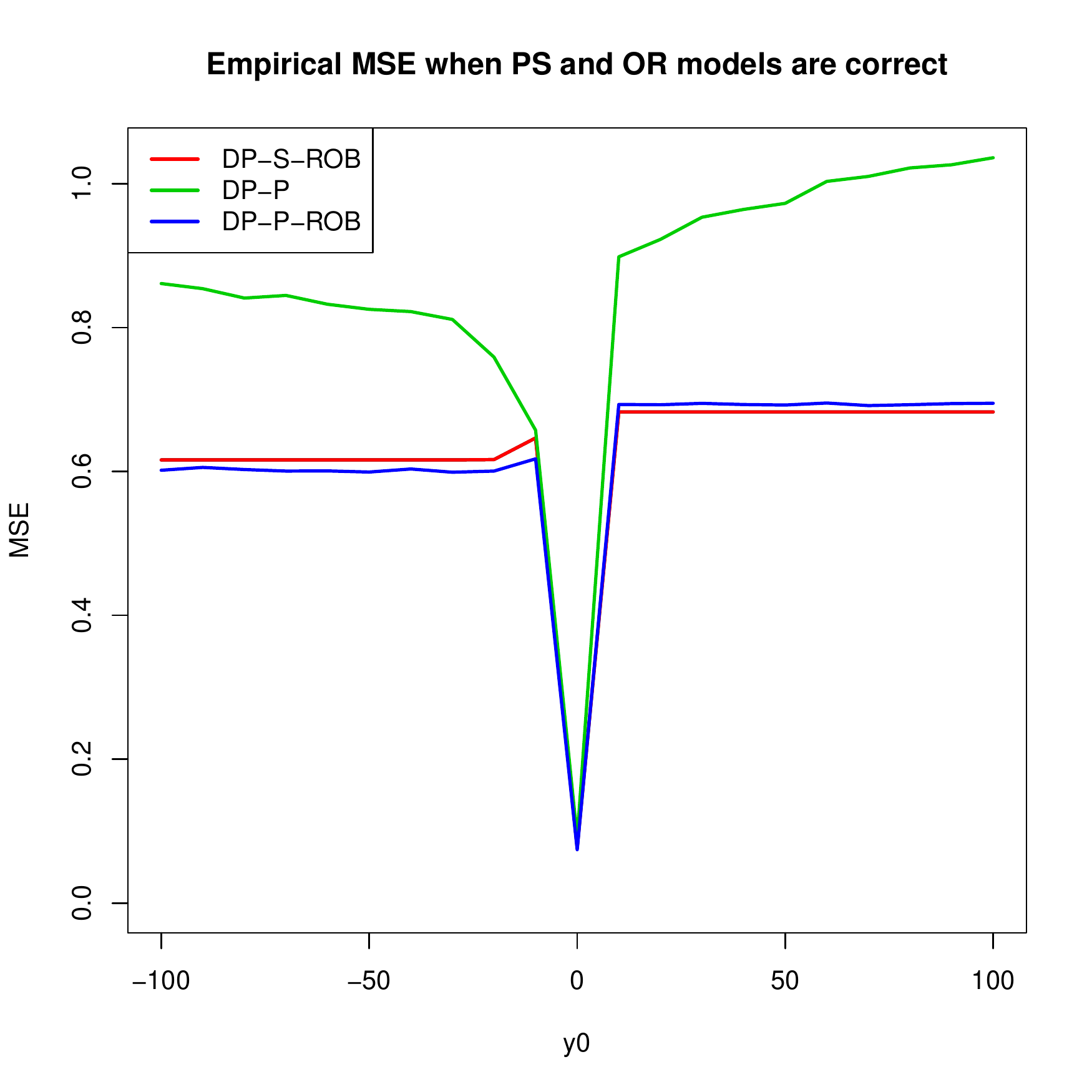}\end{center}
\caption{Doubly robust estimators in contaminated samples}
\label{fig:poro}
\end{figure}
\begin{figure}
\begin{center}\includegraphics[scale=0.5]{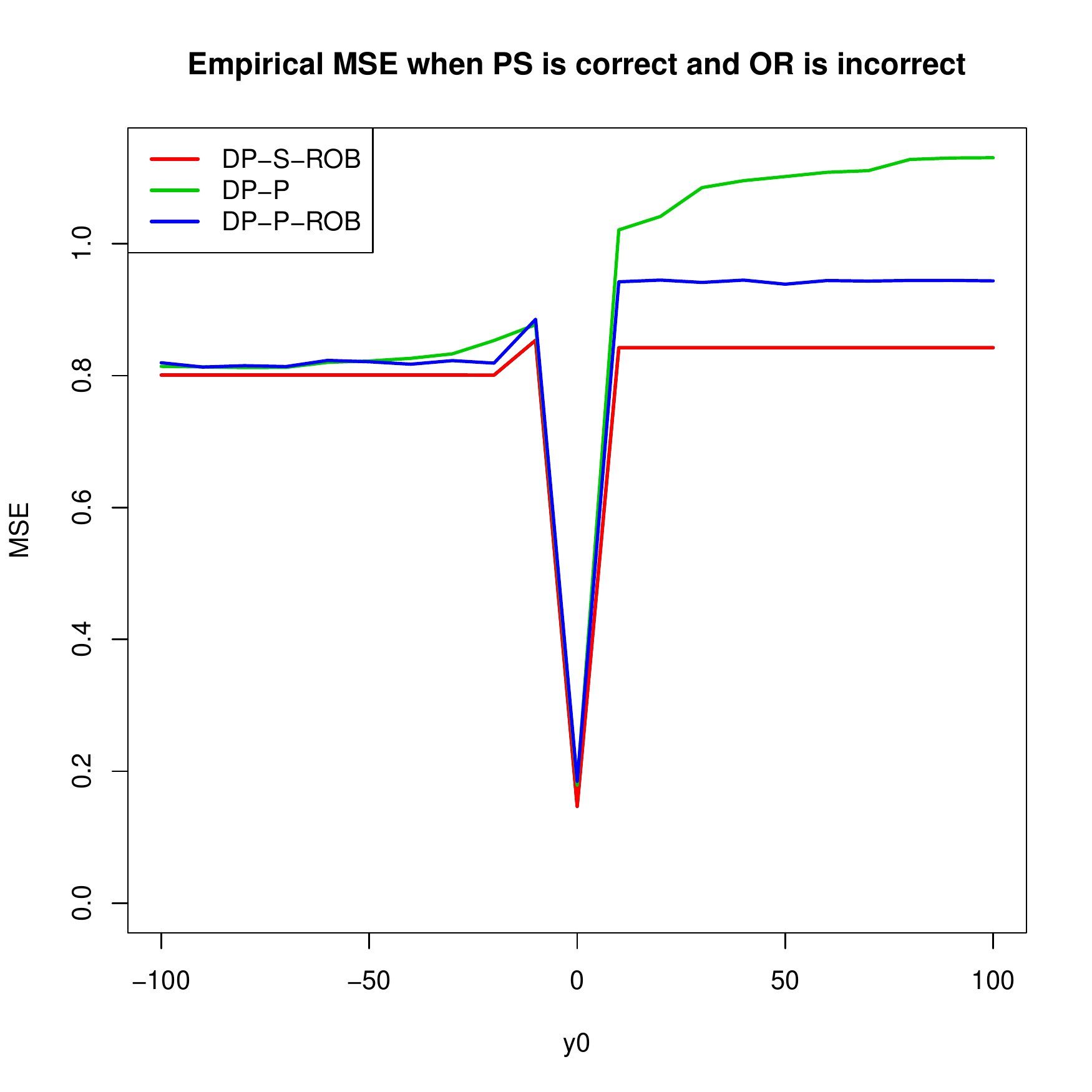}\end{center}
\label{fig:porm}
\caption{Doubly robust estimators in contaminated samples}
\end{figure}
\begin{figure}
\begin{center} \includegraphics[scale=0.5]{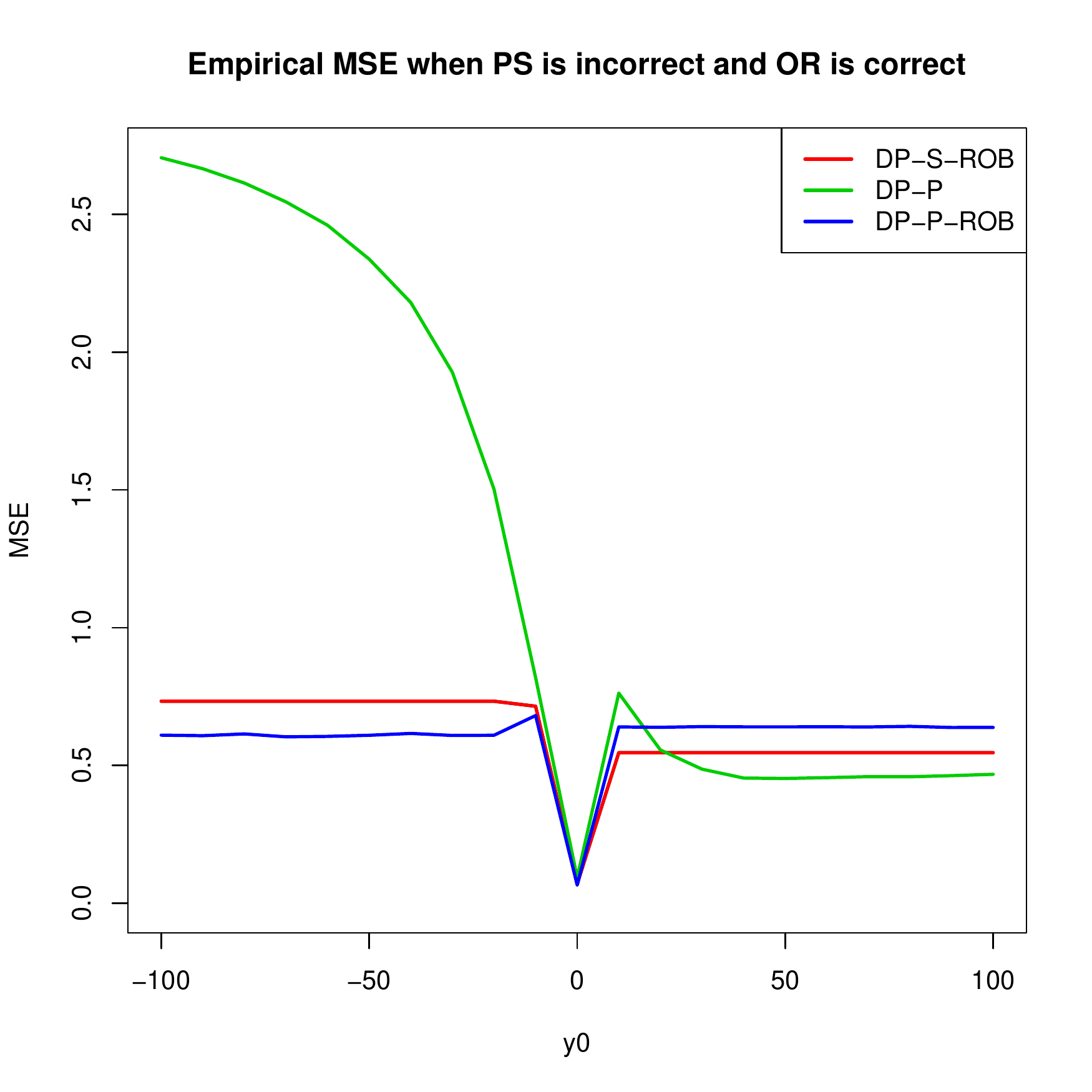}\end{center}
\caption{Doubly robust estimators in contaminated samples}
\label{fig:pmro}
\end{figure}
\begin{table}[ht]
\centering
\begin{tabular}{rrrr}
  \hline
Estimator&PS&OR&MSE\\  
  \hline
 \ipw &correct & & 0.511 \\ 
   \ipw &incorrect & & 0.392 \\ 
   \SY & &correct & 0.107 \\ 
   \SY & &incorrect & 3.117 \\ 
   \DPnorR &correct &correct & 0.146 \\ 
   \DPnorR &correct &incorrect & 0.153 \\ 
   \DPnorR &incorrect &correct & 0.127 \\ 
   \DPnorR &incorrect &incorrect & 0.155 \\ 
   \DPP &correct &correct & 0.225 \\ 
   \DPP &correct &incorrect & 0.166 \\ 
   \DPP &incorrect &correct & 0.288 \\ 
   \DPP &incorrect &incorrect & 0.164 \\ 
   \DPPR &correct &correct & 0.142 \\ 
   \DPPR &correct &incorrect & 0.158 \\ 
   \DPPR &incorrect &correct & 0.122 \\ 
   \DPPR &incorrect &incorrect & 0.160 \\ 
   \hline
\end{tabular} \caption{MSE of different estimators in different scenarios for  Hospital data. }\label{table:ex1}
\end{table}

\end{document}